# Modelling heat transfer in laser-soft matter interaction via chaotic Ikeda map

**Morteza A. Sharif[1,*], M. Shahnazary[1], M. Pourezzat[1]**

Optics & Laser engineering group, Faculty of Electrical engineering, Urmia University of Technology, 5716617165, Band road, Urmia, Iran

**Email address: m.abdolahisharif@ee.uut.ac.ir*

**Abstract-** We develop a model for simulating the heat transfer phenomena within a biological soft material using the Ikeda chaotic map. Our approach is implemented by sampling the optical intensity via the Ikeda map to investigate the influence on the heat distribution over the tissue. Our method has many potential advantages including the possibility of investigating the nonlinear optical effects resulting from the intense beam-induced feedback mechanisms. This in turn, leads to the flexibility and dynamical controllability in comparison to the quasi-static Monte-Carlo method. The proposed approach is thus appropriate for the applications in the light beam-guided nanodrug injection and microsurgery.

# 1 Introduction

Since the last two decades, the investigations on laser-tissue interaction has been accelerated particularly in biomedical optics. Understanding the mechanism of interaction and radiation/heat transfer can truly delineate a proportional strategy for diagnosis, treatment and the procedure of drug delivery [1-11]. Meanwhile, laser radiation imposes its influence by the various mechanisms including the reflection, refraction, absorption and scattering, each of which depending on the exposure time, tissue structure and radiation wavelength and power [12-14]. The influence appears through the different phenomena like the photodisruption, ablation, hyperthermia, optical damage, etc [15-21].

Monte-Carlo method is an efficient technique for simulating the linear and nonlinear statistical and quasi-static processes. It has been firstly invented in 1948 and rapidly used in different scientific areas. It has been also considered as a universal tool for developing a random-number generation algorithm required for simulating the radiation/heat transfer within the biological tissue [22-27]. However, the interaction of the laser beam with a biological and soft matter is of a nonlinear optical type and thus, an alternative algorithm should be developed. On the side, biological soft materials refer to the colloidal suspensions coupled with biological tissues [28]. For example, blood fluid is composed of suspended blood cells flowing in the blood vessel. The latter can also convey suspended micro/nano drug toward the aimed organism. This can be achieved by a beam-guiding process promising for the cancer therapy in nanomedicine [1]. Generally, biological soft materials have rheological properties similar to the colloidal suspensions and their shape and physicochemical characteristics can be modelled by a complex fluid [28]. Meanwhile, the nonlinear interaction due to a strong beam intensity can modify the physical properties of the biological soft materials as well as inducing a temperature gradient within the tissue.

Chaos is a dynamical feature of a nonlinear system which deals with a feedback procedure. In accordance with chaos theory, a small change in the initial conditions can lead to a large variation in the system dynamical state. In contrary to the randomness, chaotic regime follows a deterministic map. In this connection, Ikeda map is a universal model to describe the route to chaos in nonlinear optical systems which appears in the form of unpredictable complex behaviour [29-33].

In this study, we use the Ikeda map for developing an algorithm required for simulating the influence of the intense laser beam on the heat transfer phenomena in a biological soft material. Although the Monte-Carlo-based algorithms can model the nonlinear processes, Ikeda-based algorithm is adaptive more to the nonlinear optical phenomena since it does not follow a random procedure but imitates a dynamical pattern. Ikeda map can also represent the states prior to the instability and chaos within the form of periodic or quasi-periodic behaviour. This can be implemented by adjusting the control parameters. Therefore, a chaos-based algorithm can model the processes like the light beam-guided microsurgery, drug delivery and injection through which controlling the diffusion/scattering rate of the radiation/heat transfer is crucial. The proposed algorithm model can be also combined with the Monte-Carlo method. It can be also adapted to the condition of which the tissue structure/material suddenly changes and therefore, the underestimation/overestimation can be correspondingly modified.

## 2 Theory

The procedure of heat transfer in a tissue induced by a laser radiation can be described by Eq.(1) [13-14].

$$\nabla^2 T(r,t) = \frac{1}{\kappa}\left(S(r,t) - \rho c \frac{\partial T(r,z)}{\partial t}\right), \tag{1}$$

where $T$ is the temperature; $\kappa$ is the tissue thermal conductivity; $\rho$ is the density; $c$ is the specific heat and $S$ is the laser power transferred per volume unit. The latter can be written as $S = (\mu_s + \mu_a)I$ whereas $I$ is the laser beam intensity and $\mu_s$ and $\mu_a$ are respectively the scattering and absorption coefficients of the tissue. Writing $\mathbf{q} = -\kappa \nabla T$, one can obtain Eq.(2) as the integral form of Eq.(1) [4].

$$\int \mathbf{q}.\mathbf{n}dA = \rho c \left( \int \left( \frac{\partial T}{\partial t} - S \right) dv \right). \tag{2}$$

Eq.(1) can be discretised using the finite difference method as given in Eq.(3).

$$\frac{T_{i-1,j,k}^n - 2T_{i,j,k}^n + T_{i+1,j,k}^n}{(\Delta x)^2} + \frac{T_{i,j-1,k}^n - 2T_{i,j,k}^n + T_{i,j+1,k}^n}{(\Delta y)^2} + \frac{T_{i,j,k-1}^n - 2T_{i,j,k}^n + T_{i,j,k+1}^n}{(\Delta z)^2}$$
$$= \frac{S_{i,j,k}^n}{\kappa} - \frac{\rho c}{\kappa} \frac{T_{i,j}^{n+1} - T_{i,j}^n}{\Delta t}, \tag{3}$$

where $i$, $j$ and $k$ are respectively the numerical counter for the spatial coordinates $x$, $y$ and $z$; $n$ is the same for time; $\Delta x$, $\Delta y$, $\Delta z$ and $\Delta t$ are respectively the spatial and temporal step sizes. Eq.(1) has been extensively investigated in various studies through which the Monte-Carlo algorithm has been considered as the principal idea to simulate the temperature gradient within the tissue [22-27]. In contrast, complex chaotic map of Ikeda instability is used in this study to simulate the laser intensity-induced thermal pattern within the tissue. The main reason behind this idea is the observation of self-focusing and subsequently, the strong nonlinear response in biological soft materials triggered by an intense beam intensity [1,33-35]. The nonlinear response is anticipated to be further enhanced during a light beam-guided microsurgery, drug delivery or injection in which the colloidal nanoparticles used for remedial purposes are attracted toward the zones with the higher intensity and then, providing a large nonlinear refraction [1,33-35]. The procedure of applying the Ikeda instability during the simulation stage is independent from implementing the Monte-Carlo algorithm in which the linear scattering/absorption coefficient is to be modified. Here, the intensity of the light beam $I$ is modelled in accordance with the Ikeda chaotic map given in Eq.(4) [33].

$$\left( I_{i,j,k}^{n+1} \right)^{1/2} = \left( I_0^n + \varsigma \left( I_{i,j,k}^n \right)^{1/2} \exp\left[ i \left( \varphi_0 + I_{i,j,k}^n \right) \right] \right),$$
$$\left( I_{i,j,k+1}^{n+1} \right)^{1/2} = \left( I_{i,j,k}^{n+1} \right)^{1/2} \exp\left[ -\left( \mu_a + \mu_s \right) \Delta z \right], \tag{4}$$

where $I_0^n$ is the input intensity; $\varsigma$ is assumed as the feedback strength lying between 0 and 1. $\varphi_0$ is the initial light beam phase. On this base, $S_{i,j,k}^n$ in Eq.(3) can be substituted with $(\mu_s + \mu_a)I_{i,j,k}^n$. As it is clear, both the Monte-Carlo algorithm and Ikeda map are simultaneously applicable.

For the modest values of $\varsigma$, the nonlinear dynamical behaviour emerges through the quasi-periodic oscillations. Otherwise, if $\varsigma$ tends to 1, a transition to chaotic regime will be resulted. $\varsigma$ is particularly modest for the nonlinear diagnostic processes like the two-photon imaging technique, coherent Raman spectroscopy, etc. through which the deep penetration of light beam can cause the higher accuracy, resolution and sensing [1]. On the other hand, in the processes like the laser-assisted nanosclale drug delivery or laser injection, nanoparticles are injected into the tissue for the treatment purpose. This is turn, leads to the local field enhancement over the surface of nanoparticles as aforementioned and then provides a strong feedback entity measured by $\varsigma$ [1,6,34]. Therefore, laser-induced temperature gradient should be precisely monitored in the above-mentioned effects to control the heat transfer phenomena and prevent from any optical damage.

## 3 Simulation results and Discussion

### 3-1 Ikeda map

As aforementioned, the map of Eq. (4) is extremely sensitive to the value of $\varsigma$. Fig.1(a) shows the temporal evolution of breather oscillations obtained for $\varsigma = 0.633$. For much longer times, the amplitude of fluctuations exponentially grows implying then, a transition to chaotic regime. On the other hand, Fig.1(b) shows the phase portrait of Ikeda map.

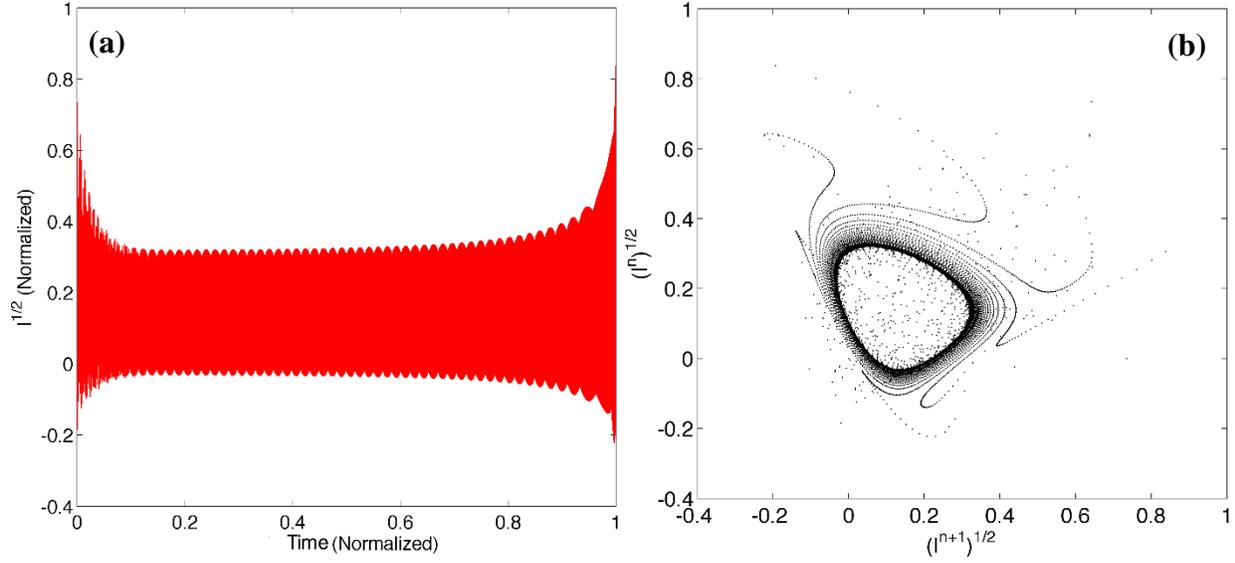

Fig.1 (a) breather oscillatory behaviour, (b) phase portrait of Ikeda map, for $\varsigma=0.633$

**3-2 Temperature gradient**

In the biological soft material, the value of $\varsigma$ remains the modest values. In consequence of the interaction of the laser intensity with tissue, the nonlinear dynamical state will thus stand away the complex chaotic regime. However, the quasi-periodic behaviour will be still dominant. In this study, we thus assume the feedback strength within the range of the modest values i.e. $0 < \varsigma < 0.25$. Presumed values for the other parameters in the simulation stage are $\mu_a = 0.6 \text{ cm}^{-1}$, $\mu_s = 100 \text{ cm}^{-1}$, $c = 3662 \text{ Jkg}^{-1}\text{K}^{-1}$, $T_0 = 37 \,^\circ\text{c}$, $\rho = 1060 \text{ kgm}^{-3}$, $k = 0.512 \text{ Wm}^{-1}\text{K}^{-1}$. A laser diode with the wavelength $850 \text{ nm}$ and power of $4.6 \text{ W}$ is considered. Cylindrical coordinate system is assumed for solving Eq.(1) numerically. The penetration depth $z = 2 \text{ mm}$ is scaled to 1 for all obtained results shown in Fig.2 for which four different values of $\varsigma$ is assumed after a two minutes irradiation. For the case $\varsigma = 0$, if the Monte-Carlo algorithm is lonely implemented, the simulation result will convey a uniform temperature pattern (Fig.2(a)). Nevertheless, for the values of feedback strength $\varsigma > 0$, temperature pattern considerably changes; the quasi-oscillatory behaviour appears as a result of the Ikeda map modulation instability. For the smaller value $\varsigma = 0.05$, the pattern – shown in Fig.2(b) – reveals a slight difference in the temperature compared to the non-feedback case (Fig.2(a)). The discrepancy becomes more evident for the pattern, obtained for the larger value $\varsigma = 0.1$ (Fig.2(c)) in which the temperature seems to penetrate more in the depth. This particularly occurs for the higher temperature zones. Further increase in the value of feedback strength ($\varsigma = 0.25$) apparently leads to the less penetration depth for the lower temperature zones, but more depth for the hot spots. This is shown in Fig.2(d) for which the periodic behaviour is also more recognizable.

As the next step, we investigate the time evolution of heat distribution in the tissue. All parameters' values are unchanged while the maximum radiation time is assumed this time to be 3 minutes. Fig.3 shows the time evolution of the temperature variation vs. the penetration depth for different values considered for feedback strength. The penetration depth for 3 minutes radiation can increase up to $0.5\,\text{cm}$. For the non-feedback case (shown in Fig.3(a)), a rather linear increase is deduced for the temperature gradient pattern during the radiation time. In the presence of the feedback entity, the linear feature is more or less vanished even for the smaller value of feedback strength ($\varsigma = 0.05$ for which the result is shown in Fig.3(b)) and is aggravated for the larger value $\varsigma = 0.1$ for which the result is shown in Fig.3(c) indicating also a shift in the depth for the hot spots as previously deduced from Fig.2(c). However, the result obtained for $\varsigma = 0.25$ (Fig.3(d)) reveals that the tissue remains with modest temperature at initial times of the radiation. This implies that the interaction with the feedback procedure can lead to the nonlinear temperature gradient within the tissue in such a degree that conducts the heat into the deeper zones.

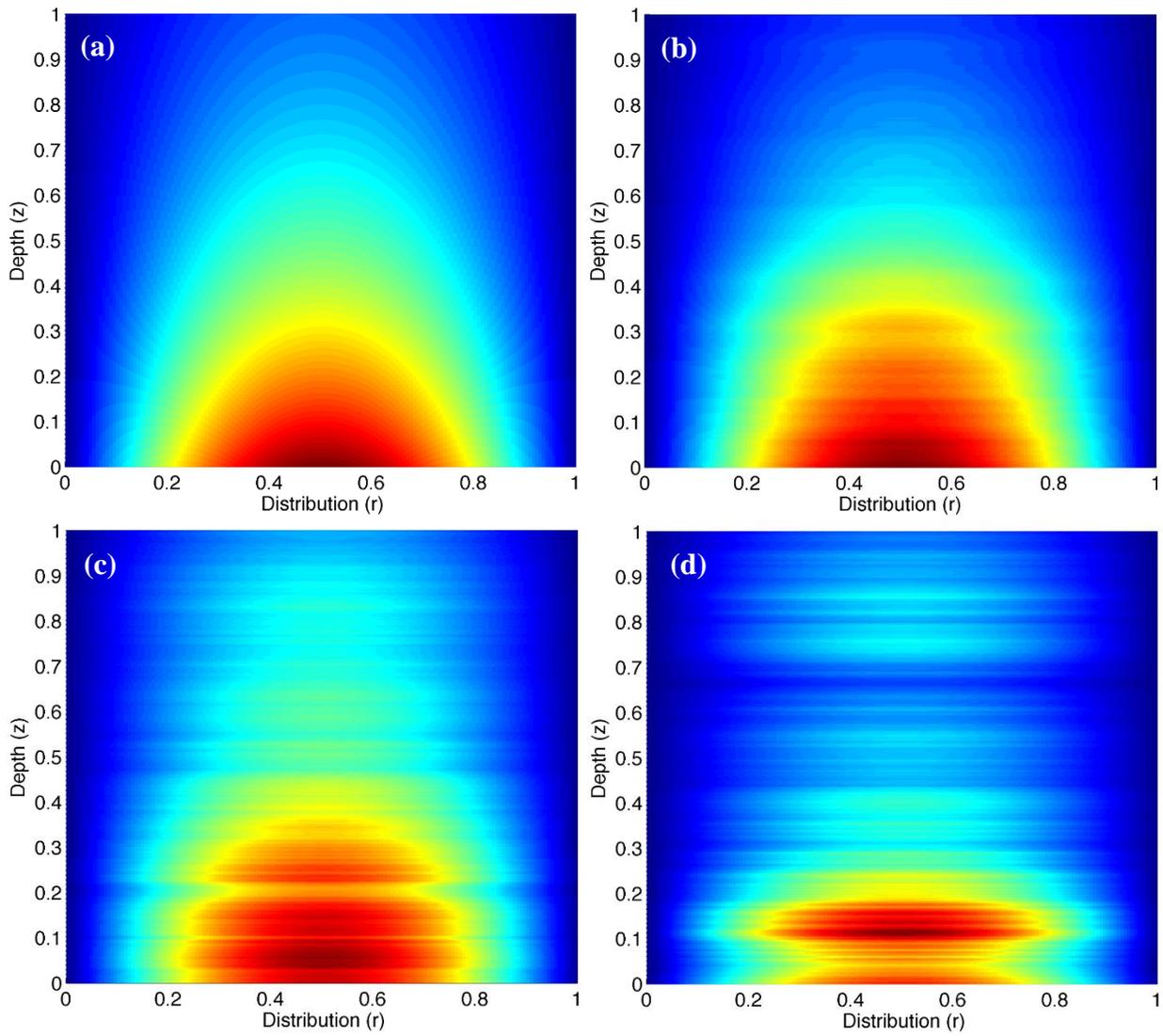

Fig.2 Temperature distribution shown in the cylindrical coordinate for different feedback strengths (a) $\varsigma = 0$ (non-feedback case) (b) $\varsigma = 0.05$, (c) $\varsigma = 0.1$, (d) $\varsigma = 0.25$.

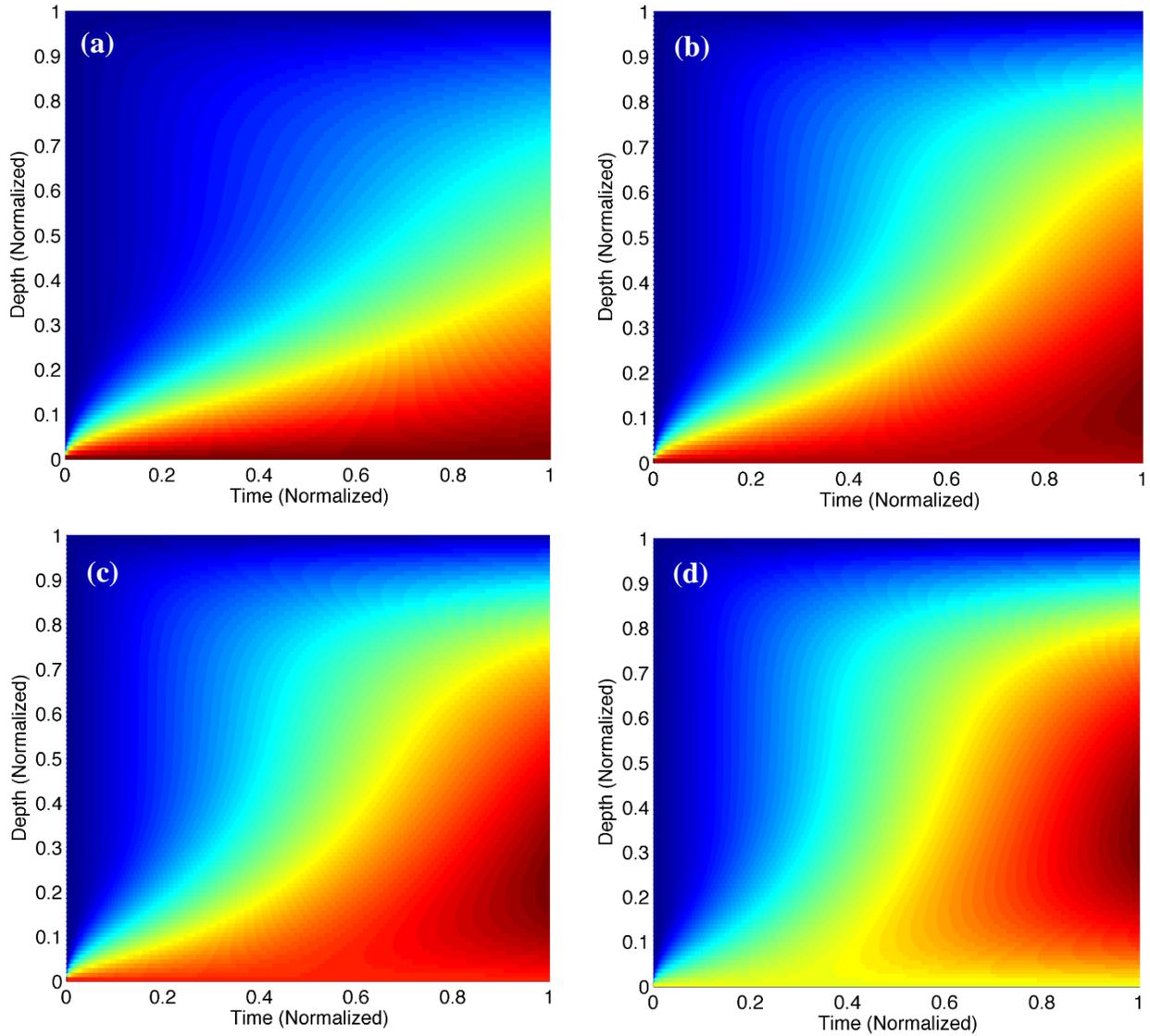

Fig.3 Time evolution of temperature variation shown in the cylindrical coordinate for $\mu_a$=-0.6 cm$^{-1}$ and for different feedback strengths (a) $\varsigma=0$ (non-feedback case) (b) $\varsigma=0.05$, (c) $\varsigma=0.1$, (d) $\varsigma=0.25$.

The effect of absorption coefficient may be also determinative. To investigate it, we assume that the absorption coefficient is increased up to two times i.e. $\mu_a = 1.2$ cm$^{-1}$. The results are shown in Fig.4. In comparison with Fig.3, it is evident that the feedback entity has less influence on heat transfer in the presence of the larger absorption coefficient.

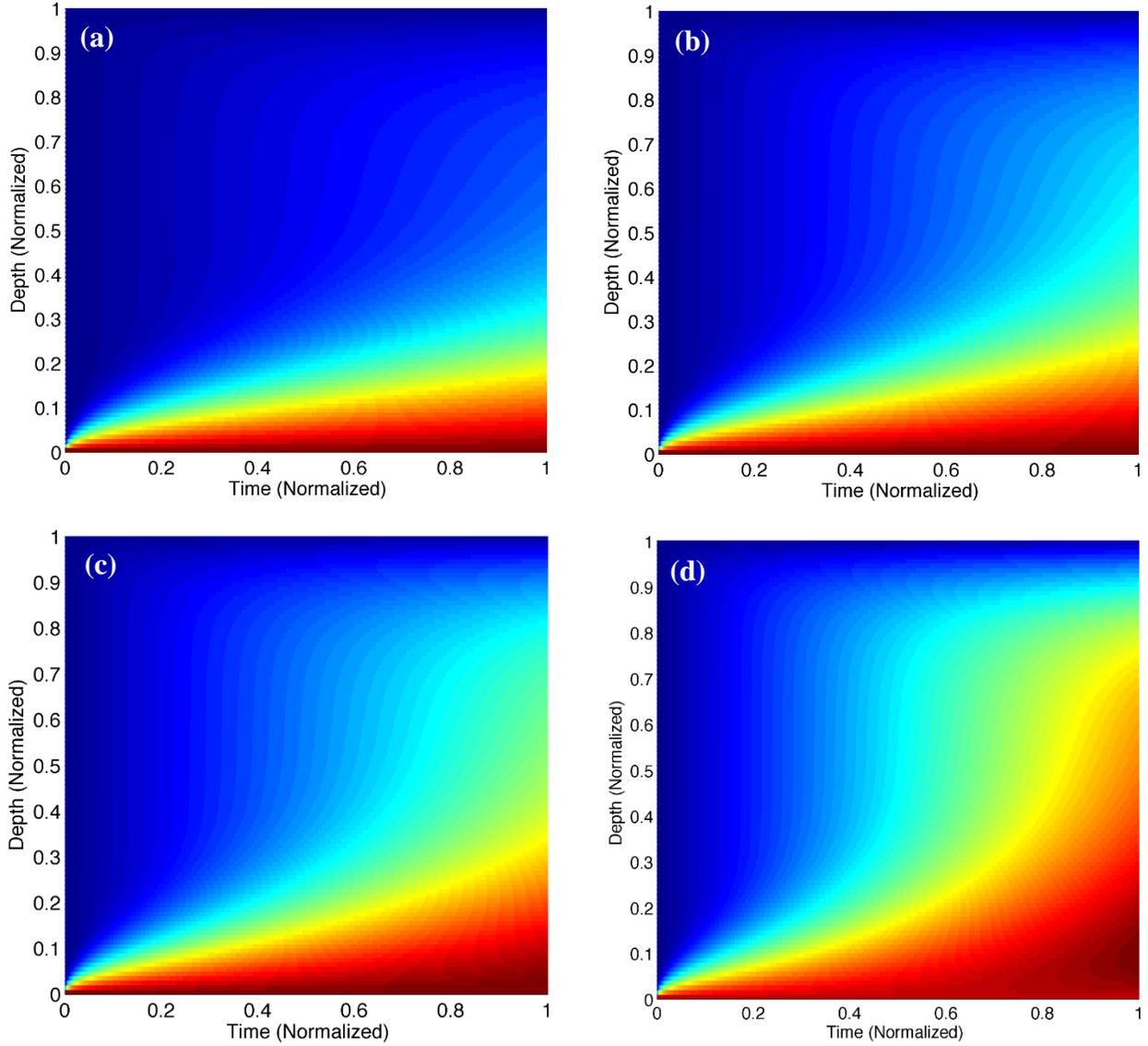

Fig.4 Time evolution of temperature variation shown in the cylindrical coordinate for $\mu_a$=-1.2 cm$^{-1}$, also for different feedback strengths of (a) $\varsigma=0$ (non-feedback case) (b) $\varsigma=0.05$, (c) $\varsigma=0.1$, (d) $\varsigma=0.25$.

One may argue about any change in the other optical parameters and particularly the nonlinear characteristics of tissue including the nonlinear refraction and nonlinear absorption. This may follow an alternative formalism to that assumed in thus study for investigating the light beam propagation within the tissue. However, there will be indeed no such alteration in a normal tissue. Although it is truly expected to have a change in the values of the optical parameters in some special process like the laser beam-guided nanodrug delivery or dye-enhanced tissue [4], since the penetration depth mostly remains trivial in comparison to the nonlinear effective length, the proposed map of Eq.(4) will be still adequate to introduce an initial nonlinear feature affecting on the heat transfer phenomena.

The model proposed in this study is flexible and has a dynamical feature. This arises from the fact that for the modest values of the feedback strength, Ikeda map shows a quasi-periodic behavior which is predictable mathematically while for the larger values of the feedback depth, an unpredictable state and a route to chaotic regime will be anticipated. As a prospect, the model is open for tany experimental verification.

**4 Conclusion**

We have proposed a modelling procedure of the laser-induced heat transfer phenomena within a tissue based on chaotic Ikeda map. Our approach is alternative to the Monte-Carlo method and has the advantages of flexibility, predictability and providing the investigation of nonlinear effects in consequence of the strong laser-tissue interaction. Hence, it is appropriate for the applications in laser beam-guided nanodrug delivery, microsurgery, etc. We have shown that the model modifies the temperature pattern within the tissue in a more sensitive manner compared to the Monte-Carlo algorithm. This arises from the fact that the action depth of our algorithm is controllable. We ascribe this modification to the laser intensity-induced nonlinearity which in turn provides a feedback procedure and thus, is different from the considerations based on the Monte-Carlo algorithm.

**Conflict of interest**

The authors declare that they have no conflict of interest.